# Nanoscopic Filters as the Origin of d Wave Energy Gaps

J. C. Phillips

Dept. of Physics and Astronomy,

Rutgers University, Piscataway, N. J., 08854-8019

ABSTRACT

A ferroelastic nanoscopic checkerboard pattern of pseudo- and superconductive gaps has been resolved by scanning tunneling microscopy on BSCCO. Using this pattern one can *derive* [not assume] a macroscopic anisotropic d wave superconductive energy gap that agrees well with angle-resolved photoemission and Fourier transform scanning tunneling microscopy data. The derivation is orbital only [no spins], it explains chemical trends in Raman scattering peak positions and strengths, and it predicts an unexpected reversal of anisotropic Raman gap signatures between BSCCO and LSCO, in good agreement with experiment.

## 1. Introduction

Recent high-resolution **r**-space tunneling (Pan et al 2001; Lang *et al.* 2002) and **k**-space photoemission (Bogdanov *et al* 2000; Johnson *et al.* 2001; Lanzara *et al.* 2001; Gromko *et al.* 2002) experiments on the spatial and momentum surface electronic structure of ceramic cuprate high-temperature superconductors (HTSC) have revealed very irregular and anisotropic features with spectral and temperature dependencies closely related to the superconductive energy gap. The most recent high-resolution Fourier transform STM experiments (McElroy *et al.* 2002) encompass an awesome data base: a 65 nm field of view encompassing 40,000 surface unit cells is scanned with 0.13 nm resolution at energy intervals of 2 meV between the limits ± 30 meV. They show that on this macro(meso)scopic length scale the planar superconductive energy gap is anisotropic with quasi-tetragonal d-wave ($x^2 - y^2$) **k**-space symmetry in very good quantitative agreement with that observed by **k**-space photoemission. This strong anisotropy, with nodes in $(\pi,\pi)$ directions, and antinodes in $(\pi,0)$ directions, contrasts strikingly with the largely s-wave gaps that are characteristic of tetragonal metallic superconductors such as Sn. The explanation for this anisotropy may well lead to an understanding of the microscopic mechanisms responsible for high temperature superconductivity.

There are at least three length scales to be considered in these spatially inhomogeneous materials: the microscopic length of 0. 3 nm (unit cell), the macroscopic length scale of optical beams and Fourier transform STM experiments, and the intermediate nanodomain 3 nm length scale of the polygonal (idealized as checkerboard) (superconductive and pseudo-) gap structure (Pan et al 2001; Lang *et al.* 2002; Phillips and Jung 2001). At first one might suppose that the explanation for d waves could be found at the microscopic level, as both conventional phonon and exotic magnon pairing interactions are short range (~ 0.3 nm). However, by themselves the phonon interactions are sufficiently isotropic that one could expect from them only a weak admixture of d wave into s wave anisotropy. The magnon interactions appear to be more promising, because antiferromagnetic order is associated with $(\pi,\pi)$ magnon scattering that may connect $(\pi,0)$ and $(0, \pi)$ gap maxima (Ruvalds *et al*. 1995; Demler and Zhang 1995; Onufrieva *et al*. 1996; Campuzano *et al*. 1999). However, apart from magnon scattering being incompatible with Cooper pairing, its composition dependence cannot explain an optimal doping far removed from the insulating antiferromagnetic region of the phase diagram where it is strongest. It appears that $(\pi,\pi)$ broadening of $(\pi,0)$ states by $(\pi,\pi)$ magnon scattering, while possibly appreciable in underdoped samples, has practically disappeared in overdoped samples where $T_c$ is still high (Bogdanov *et al*. 2002). This is just what one would expect if the magnon scattering effects were irrelevant to Cooper pair formation and occurred only at interfaces between magnon nanodomains and superconductive nanodomains. With increasing doping, the spatial areas of the former (latter) shrink (grow), and in overdoped samples the interfacial areas shrink. If scattering at such nanoscopic interfaces determined $T_c$, then $T_c$ would drop monotonically from underdoped to overdoped samples, but this is not observed. (In any case, the most that can be said of (π,π) magnon scattering is that it may help to explain the weakness of $q_4$ scattering in octet formation (McElroy *et al.* 2002; Phillips 2003)).

A possible microscopic way to derive bulk d waves from interactions between holes and phonons bound to dopants has been suggested (Phillips 2003). This mechanism relates the (π,0) gap maximum to the anisotropy of the LO phonon dispersive pseudogap (Phillips and Jung 2002), which is large in (π,0) directions and small (or zero) in (π,π) directions. This mechanism involves dopant short-range order and has many attractive features, including a strong dependence on screening that is consistent with the phase diagram (Phillips 2002a,b). Its importance has increased in the light of recent data on many (pseudo-) perovskites (Braden *et al.* 2002) that show that LO phonon anisotropy is cuprate HTSC-specific. Moreover, it may be related to pseudogap nanodomain formation, as an extra LO branch is sometimes observed, for instance, in $Ba_{1-x}K_xBiO_3$. Angle-integrated photoemission on $Ba_{1-x}K_xBiO_3$ has shown (Chainani *et al.* 2001; Yokoya *et al.* 2002) both a 5 meV superconductive gap and a 70 meV pseudogap for x = 0.33. The latter is similar in magnitude to the 60 meV pseudogap seen by STM in BSCCO. In spite of these successes, in this paper a different nanoscopic domain mechanism that also gives d waves is proposed. The question of which mechanism, microscopic bulk dopant or planar nanodomain filter, is correct, has important implications for all theories of high temperature superconductivity. After describing the nanoscopic mechanism in the next Section, experimental evidence from Raman scattering is described in Section 3 that bears on this question.

## 2. Nanodomain Interfacial Filters

The checkerboard structure observed by STM (Pan et al 2001; Lang *et al.* 2002) is irregular, and it consists of alternating "squares" with narrow $\Delta_s \sim 40$ meV (superconductive) gaps and broad $\Delta_p \sim 60$ meV pseudogaps. The superconductive order parameter $\Delta_s$ is the Cooper pair amplitude ψψ. At the nanoscopic level one supposes that ψ forms octets (McElroy *et al.* 2002) because these are Wannier functions for dopant states near the Fermi level in each superconductive "square" nanodomain (Phillips 2003). It is natural to assume that the polygonal edges of the idealized square are on the average parallel to the "dense" a and b axes of the $CuO_2$ planes that contain Cu-O-Cu bonds, and that Fermi energy electrons can be reflected by Jahn-Teller distorted Cu-O-Cu rows. [This assumption is similar to reflection of surface electrons by steps on Au and Cu, as observed by Fourier transform STM (Petersen *et al.* 1998). Reflections can also occur from impurities, and these can be used to map out surface bands (Song *et al.* 2001; Morgenstern *et al.* 2002), but such reflections do not generate new anisotropies).] In BSCCO there are two Cu atoms in each planar unit cell, Fig. XI.2 of (Phillips 1989), which causes the real-space **r** axes [a,b] to be rotated by $\pi/4$ relative to the **k**-space axes (x,y). However, in LSCO there is just one Cu atom in each planar unit cell, Fig. III.12 of (Phillips 1989), and the [a,b] and (x,y) axes are parallel.

The d-wave gap measured in optical experiments and by scanning tunneling spectroscopy necessarily involves many nanodomains. (In fact, the coherence length of the octet wave packets is just 1.5 nm, the average nanodomain radius.) Thus to define the gap $\Delta_s$ one must percolatively connect the amplitudes of $\Delta_s$ across the interfaces between many superconductive squares (Phillips and Jung 2001). The most natural way to do this, in the presence of strong disorder, is to assume that the normal components of $\mathbf{v}_s = \nabla \ln(\psi\psi)$ are continuous across the interface. For BSCCO the [11] spatial normals **n** diagonally connecting superconductive "squares" are rotated by $\pi/4$ to be parallel to $(\pi,0)$ directions in **k** space. Then the normal components are

$$\Delta_{ns} = \Delta_{0s} \exp(2i\theta) \qquad (1)$$

where $\theta$ is the angle between **k** and $(\pi,0)$. The measured gap parameter is the real part of $\Delta_{ns}$, which is $\Delta_{0s}\cos 2\theta = \Delta_{0s} (k_x^2 - k_y^2)/(k_x^2 + k_y^2)$. This is the "d wave gap"; it is the percolatively filtered residue of the Wannier nanodomain surface gap $\Delta_{0s}$.

### 3. A Bulk Experiment: Raman Scattering: BSCCO and LSCO

Mueller (2002) has emphasized that almost all the evidence for d wave gaps is based on surface experiments, such as photoemission and tunneling, while experiments that probe the bulk of the sample (such as relaxation of femtosecond infrared pulses) show little evidence of d-wave sub-gap states. Here we have introduced a second distinction, between macroscopic, nanoscopic, and microscopic length scales. Relaxation of femtosecond infrared pulses (Kabanov and Mihailovic 2002) is not only a bulk experiment, but it also probes microscopic length scales. Raman scattering is an interesting intermediate case, as it can probe the bulk. It is both macroscopic (long light wave length) and nanoscopic, as it is sensitive to nanoscopic phase separation (Bhalla *et al.* 1999, Boolchand and Bresser 2000), but it cannot determine nanoscopic dimensions. The two configurations of polarization of incident and scattered light, $B_{1g}(xy)$ and $B_{2g}((x + y)(x - y))$, scatter between the $[(\pi,0)$ and $(0,\pi)]$ and $[(\pi,\pi)$ and $(\pi,-\pi)]$ regions of **k** space, respectively (Devereaux *et al.* 1994). For homogeneous d wave anisotropy $2\Delta_{ns} = 2\Delta_{0s}\cos2\theta = 2\Delta_{0s} (k_x^2 - k_y^2)/(k_x^2 + k_y^2)$ these two configurations should give sharp peaks at $2\Delta_{0s}$ and a broad peak at $1.4\Delta_{0s}$, respectively.

For $Bi_2Sr_2Ca_{1-x}Y_xCu_2O_{8+\delta}$ at $x = 0$ (BSCCO) one observes (see Fig.1(a,b)) peaks near 350 cm$^{-1}$, nearly ten times stronger for $B_{1g}$ than for $B_{2g}$ polarization (Sugai and Hosokawa 2000). Because of its lack of nanoscopic resolution, only one Raman peak is observed at a given composition. [This appears to be a result of complex local field dynamical interactions.] Broadly speaking, the observed $A_{1g}$ and $B_{1g}$ gaps shift from ~ 40 meV at $x = 0$ to ~ 100 meV, underdoped, $x = 0.3$. At first sight these values do not seem to agree well with the bimodal STM idealized checkerboard values of $\Delta_s \sim 40$ and $\Delta_p \sim 60$ meV seen by STM. Nevertheless, the underdoped upper limit ~ 100 meV is easily understood: it corresponds to roughly the pseudogap $2\Delta_p$, which is reasonable, as the latter is expected to dominate the underdoped spectrum, and it may be somewhat reduced from 120 meV by Y-induced disorder.

With increasing doping, the observed peak, weighted by dynamical local fields, gradually is transformed so that it emphasizes more metallic regions. The overdoped lower limit ~ 40 meV seems to be too low, but if we suppose that at the interfaces between the pseudogap and supergap nanodomains there is some material that is merely an ungapped normal metal, $\Delta = \Delta_n = 0$ then in overdoped material one would expect to observe not $2\Delta_s$ but $\Delta_n + \Delta_s = \Delta_s$. [Note that larger gap excitations tend to be screened by smaller ones.] Of course, these gap values cannot be explained by assuming that superconductivity and d charge density waves coexist in the same space, as in continuum "flux phase" models, with or without "impurity scattering" (Zeyher and Greco 2002). [The macroscopic invariance of the pseudogap under combined translational and rotational symmetry proved in the Heisenberg "flux phase" model (Chakravarty *et al.* 2001) is incompatible with superconductivity and is flatly contradicted by the ferroelastic checkerboard pattern (Pan et al 2001).] Note also that in photoemission experiments where x is fixed at 0 and doping is varied by changing $\delta$, the changes in gap anisotropy with doping are small (Mesot *et al.* 1999).

Overall, the agreement of the Raman data with the above theory at $x = 0$ is reasonable, as are the chemical trends in the observed $A_{1g}$ and $B_{1g}$ gaps. The relative weakness of the $B_{2g}$ peak is easily explained by lack of percolation, or smaller coherence lengths, in [10] (11) compared to diagonal checkerboard [11] (10) directions (filter mechanism). Chemical trends in the $B_{2g}$ energy (peaking at optimal doping) can also be explained this way in terms of gap dependence on coherence length. However, even this kind of agreement is not obtained (see Fig.2(a,b)) for $La_{2-x}Sr_xCuO_4$ ($x = 0.1$) (Venturini *et al.* 2002). Now a strong Raman peak is observed only for $B_{2g}$ polarization, with a far infrared anomaly below 100 cm$^{-1}$ for $B_{1g}$ polarization. (The latter is interpreted by the authors as evidence for a sliding (weakly pinned) charge density wave, an interpretation supported by their infrared data (not illustrated here)). No pair-breaking Raman peak is observed for $B_{1g}$ polarization; the authors explain this absence as a result of underdoping, analogously to $Bi_2Sr_2Ca_{1-x}Y_xCu_2O_{8+\delta}$ at $x = 0.2$, where, however, the peak is shifted to lower energy but is still observable. In the d-wave context these observations, and in particular the persistence of the $B_{2g}$ peak, although the $B_{1g}$ peak disappears with underdoping, is surprising, as both the gap and the node are observable by ARPES in underdoped samples where $x = 0$ and only $\delta$ is varied (Mesot *et al.* 1999).

The percolative nanoscopic filter mechanism discussed above suggests a different explanation for the $La_{2-x}Sr_xCuO_4$ ($x = 0.1$) Raman anomaly. Because there is only one Cu/unit cell, in this case the real-space **r** axes [a,b] are not rotated by $\pi/4$ relative to the **k**- space axes (x,y). So long as the percolative filter in the checkerboard remains diagonally oriented relative to the real-space **r** axes [a,b], this means that the anisotropic d-wave gap should have its maximum now in the (1,1) direction, not the (1,0) direction, as in BSCCO. That gap has so far not been resolved by ARPES, presumably because of difficulties in surface preparation (Ino *et al.* 2002). However, one can say that the Raman pair-breaking LSCO $B_{2g}$ peak in Fig. 2(b) could be associated with the (1,1) gap maximum predicted by the percolative filter model.

The alternative projective d-wave gap mechanism suggested earlier (Phillips 2003) based on strong coupling of carriers to LO phonons bound to filaments preferentially oriented, may still have some validity. It was noted (Phillips 2001a) that the strongly anisotropic LO phonon gap is observed not only in the cuprates, but also in $Ba_{0.6}K_{0.4}BiO_3$, which has a lower $T_c$, but otherwise is similar to the cuprates. However, this mechanism does not explain the apparent (11) orientation of the LSCO Raman anomaly.

## 4. Other Bulk Experiments: Specific Heat and Point Contact Tunneling

Higher order magnetic field corrections to macroscopic properties (specific heat, Hall and thermal resistivities) have been interpreted (Simon and Lee 1997) in terms of d-wave gap nodes. The theory assumes microscopic sample homogeneity and is not self-consistent, that is, the derived microsocopic quasi-particle states do not yield a d-wave gap. In the present model the checkerboard structure is essential to the origin of the d wave gap. However, in the macroscopic limit ($\mathbf{H} \to 0$) properties such as the specific heat will simply average over the surface areas and d-wave behavior will be observed in the corrections. Transport properties appear to follow the predicted behavior at low magnetic fields, which implies little or no scattering (mean free paths long compared to 3 nm) at percolative checkerboard interfaces. This is consistent with a single-mode filamentary model, analogous to single-mode optical fibers.

Point-contact tunneling spectroscopy has occasionally shown gap anisotropy reduced from 100% to only 3% (Shimada *et al.* 1995); with such an isotropic gap one can observe Eliashberg electron-phonon fine structure, much as with metallic superconductors, and with a localized gap amplitude $\Delta_l \sim 25$ meV. In the present model this is interpreted as evidence for tunneling into an exceptionally large and isolated nanodomain where $\psi\psi$ does not percolate to other nanodomains, and so is not subjected to interfacial filtering. This picture is self-consistent in the sense that the electron-phonon fine structure observed in the isolated case is not observed in the percolative case, because in the latter case there seems to be no damping by electron-phonon interactions at the filter. The fact that $\Delta_l \sim 25$ meV $< \Delta_0 \sim 40$ meV could be attributed to the reduction of $\Delta_l$ by enhanced Coulomb interactions in the localized case.

5. **Extension to Three Dimensions: Metal-Insulator Transition**

Nanoscopic domain structure in two dimensions can have unexpected consequences, especially when one tries to discuss data in the context either of an effective medium (homogeneous) model, or of a hybrid macroscopic-nanoscopic model based on long range "stripes". For example, (Yoshida *et al*. 2002) have reported that even for small $x < x_0 = 0.06$, an ARPES peak is present in $La_{2-x}Sr_xCuO_4$ near $E_F$ and $(\pi/2,\pi/2)$ [that is, the nodal point on the Fermi "arc"] whose intensity $I(x)$ scales roughly with x from 0 to 0.1. This is surprising because the low-temperature, low energy metal-insulator transition occurs at $x = x_0$ (however, there is a break in slope of $I(x)$ between the data points $x = 0.05$ and $x = 0.07$). In a checkerboard percolative model, of course, it is possible for the "metallic" nanodomains to contribute such an ARPES peak because the final state mean free path is comparable to the nanodomain dimensions. The intensity of that peak should scale not with the low-energy conductivity but with the area occupied by the "metallic squares". Even when the "metallic squares" occupy less than half of the planar area, irregularities in size and shape of squares will cause maximum nanodomain overlap in **r** space along [11] diagonal filters, which are parallel to the $(\pi/2,\pi/2)$ **k**-space direction for $La_{2-x}Sr_xCuO_4$ (see above). This explains the observed concentration of the peak intensity near $(\pi/2,\pi/2)$. The width of the peak, their Figs. 2(b,c), is similar to that observed for the octet structure in BSCCO, suggesting that the LSCO width is determined by nanodomains with diameters $\sim 3$ nm (correlation length $\sim 1.5$ nm). The **r** space overlap will be enhanced by filamentary correlation (better low energy dielectric screening) for x above the percolative metal-insulator transition at $x = x_0$, which also explains the observed break in slope of I(x) between x = 0.05 and x = 0.07. Finally, the peak broadens and disappears above x = 0.15, optimal doping, where the overlap between metallic nanodomains becomes so large that the angular filtering effect at nanodomain corners disappears, allowing percolation over wider angles.

This excellent agreement between the nanodomain filamentary percolation model and an ARPES anomaly is still only part of the story. The STM data on the checkerboard structure in BSCCO show that the area occupied by the "metallic squares" is about half of the planar area at optimal doping (Lang *et al.* 2002), is much less than half for underdoped samples, and presumably is more than half for overdoped samples. If conduction were entirely planar, then this would predict a metal-insulator transition at optimal doping (x = 0.15, not x = 0.06, in LSCO). The observed low energy metal-insulator transition at x = $x_0$ is what one would expect from the filamentary model, as an important part of the model (Phillips 1990) is resonant tunneling through interlayer dopants (such as Sr in LSCO) where the strong electron-phonon interactions occur that giver rise to HTSC. This enables the filaments to *bypass* semiconductive nanodomains percolatively (Phillips 1990; Phillips and Jung 2002b) when the fraction of metallic area is less than half. The model also predicted that the superconductive gap would be large in the chains of YBCO, and small or even zero in the $CuO_2$ planes; this prediction, which is contrary to the assumption of almost all other theoretical models that superconductivity occurs primarily in the $CuO_2$ planes, is in excellent agreement with recent STM observations (Derro *et al.* 2002; Misra *et al.* 2002).

The next question is to what extent this interplanar zigzag motion affects the corner [11] planar corner filtering mechanism. Because the filament passes through interlayer resonant tunneling centers and planar nanodomain corners in series, the c-axis motion has little effect on the projective effects of the filters. This explains why the d-wave character of the ARPES and Fourier STM data is nearly independent of composition. It also helps us to understand why the normal state transport anomalies occur in the intermediate phase that is host to HTSC, and most ideally at the optimal composition where $T_c$ is highest.

## 6. Conclusions

The almost universal adoption of effective medium and/or stripe models of cuprate HTSC has led to the impression that Fermi-liquid like models, with **k** as a good quantum number, provide a good description of the basic physics of HTSC (Chakravarty *et al.* 2001). For example, it is widely assumed that such models can explain the origin of square vortex lattices by giving a proper description of the vortex core, but in fact they misorient the observed lattice by π/4 (Gilardi *et al.* 2002). Similarly, it is now clear that the absence of zero-energy vortex core states is a consequence of the preference of vortices for pseudogap regions over superconductive regions (Hoogenboom *et al.* 2001; Hoffman *et al.* 2002). To the extent that the checkerboard (superconductive and pseudo) gap structure plays an important part in the microscopic physics, such translationally invariant **k**-spaced approaches cannot be correct. Indeed, it is obvious from general considerations that the presence of a pseudogap (Phillips 1987) in a strongly disordered material by itself means that **k** cannot be a good microscopic quantum number, as it does not describe the pseudogap band splitting, unless that occurs because of the macroscopically uniform formation of charge density waves pinned to $\mathbf{k}_F$. Such CDW's simply do not describe the STM checkerboard data, and cannot be used to explain the pervasive presence of pseudogaps. Even if macroscopically uniform charge density waves were present in the bulk, they would imply only superconductivity *via* freely sliding (unpinned) CDW's, but such unlikely objects do not yield a Meissner effect.

This paper has shown that these perplexing issues can be resolved by analyzing the physics not only at the microscopic and macroscopic length scales, but also specifically at the ferroelastic *nanoscopic* length scale, where *percolative filters* play a crucial role. Thus the percolative filamentary nanodomain model explains many contradictory features of the experimental data, all with the use of previously presented concepts. In particular, here the d-wave anisotropy of the macroscopic surface superconductive gap has been *derived* (not assumed) entirely from orbital considerations alone (no spins). The origin and Raman anisotropy of the pseudogap (Opel *et al.* 2000) is a much more complex question that may be discussed elsewhere.

It might appear from the present derivation of the d-wave anisotropy of the surface superconductive gap that all is well and good for traditional continuum models based on order parameters, etc. This is not correct, even apart from the fact that such models do not include the checkerboard structure, and cannot derive d waves, but must add them *post hoc* in a manner that cannot be reconciled with superconductivity, cannot explain the phase diagram, and so on. The point is that the octet states (McElroy *et al.* 2002), for example, represent no more than 10% of the states contributing to the tail of the tunneling gap, and the latter in turn represent a small fraction of all the states near $E_F$. The remaining 90% (or 99%) appear as noise when one attempts to project them on **k** and **q**. What are these non-**k** representable "dark states" near $E_F$? It seems unlikely that they are defect states, as the latter tend to be split away from $E_F$ by Jahn-Teller effects. As they are pinned to $E_F$, it is natural to think of them as filamentary states that somehow have not passed through the corner filters. This could be because they represent nanodomain boundary closed orbits [**k** = 0 or a zone boundary], while the octet states represent open orbits. One could then speculate about possible kinds of Josephson couplings between sets of open and closed orbits. At present such questions remain unsettled, but it is certain that continuum models are not even providing useful formulations of these questions or of the other microscopic questions that have actually been not only formulated, but at least partially answered above.

There are interesting optical analogies to the filter mechanism discussed here that have been realized in an electronic context. (Manoharan *et al.* 2000) have shown that multiple reflections from Kondo impurities arranged as elliptical Fano electron resonators (at $\omega = 0$) can refocus signals from a Kondo impurity inside the ellipse as a quantum mirage. This is enough to show that coherent quantum percolation along self-organized dopant filaments that have been filtered by scattering from Cu-O-Cu rows represents a very plausible microscopic mechanism for HTSC.

Going further afield, the microscopic FTSTM patterns that give rise to apparent macroscopic d wave gaps have the same kind of spotty pattern that has been observed for cosmic background radiation at 400,000 years (when the photon-baryon fluid decoupled) both in the infrared and by microwaves (Overbye 2003). The latter are interpreted quite simply in terms of the standard flat universe model, which is "amazing [considering] that we know so little about the dark matter and the dark energy that comprise 95% of it". If we suppose that the dark matter is neutrinos, then again one could suppose that this dark matter is acting only as a passive filter for the photon component of the coupled fluid. This could affect the apparent measured gravitational curvature in much the same way that the Cu-O-Cu rows filter the microscopic gaps to give them macroscopic d wave symmetry.

*Postscript.* The lately added Fig. 4 (McElroy *et al.* 2002) contains a subtle feature that confirms the essential features of the filter mechanism. Note that the checkerboard edges in Fig. 4A are at 45° to the atomic rows that are clear in the quantum diffraction patterns evident in Figs. 4B,C,D. Also note that the filter mechanism is generic and is not specific to cuprates. A gap with line nodes provides the best fit to STM gap and vortex data (Lupien *et al.* 2003) on $Sr_2RuO_4$.

There have been many efforts to predict impurity and vortex states within effective medium models using positive energy scattering theory (Hirschfeld and Atkinson 2002) and a d wave gap. These have failed in almost all areas, as shown by comparison with experiment (Hussey 2002). One cannot ignore nanodomains and self-organized filaments and still expect to achieve agreement with experiment.

# FIGURE CAPTIONS

Fig. 1. Raman difference spectra for BSCCO, corrected for thermal factors. The difference shows the effects of formation of superconductive gap, or enhancement of pseudogap, or some mixture of the two, from above $T_c$ to below $T_c$. These sketches are presented here for the reader's convenience, and are based on the much more extensive data of (Sugai and Hosokawa 2000).

Fig. 2. Raman spectra for $La_{2-x}Sr_xCuO_4$ (x = 0.1), sketched from (Venturini *et al.* 2002).

Fig. 1

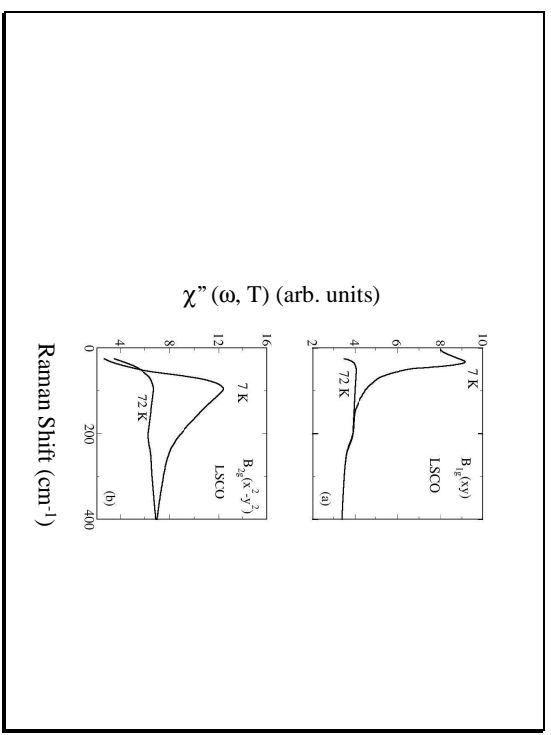

Fig. 2

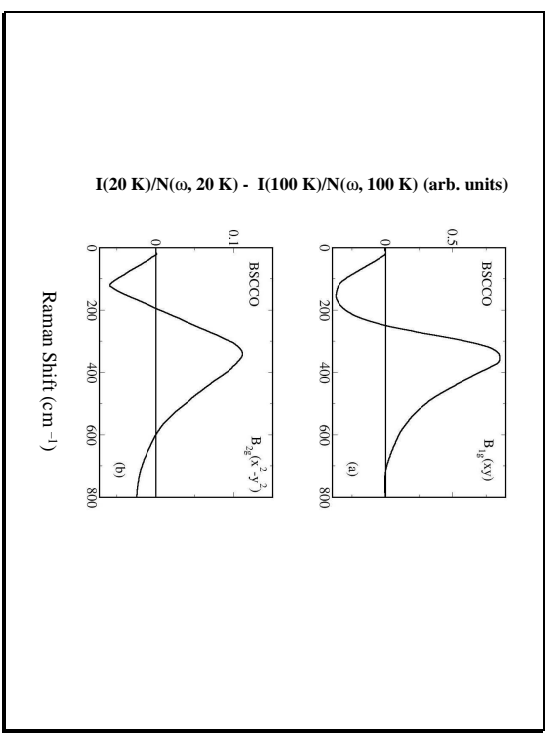

18